\documentstyle[12pt,aaspp4,epsfig]{article}

\newcommand{\grs}      {\mbox{\rm\,GRS~1758--258}}
\newcommand{\onee}     {\mbox{\rm\,1E~1740.7--2942}}

\begin{document}
\input epsf

\title{Long-Term X-ray Monitoring of 1E~1740.7--2942 and GRS~1758--258}

\author{D. S. Main\altaffilmark{1}, D. M. Smith\altaffilmark{1},
W. A. Heindl\altaffilmark{2}, J. Swank\altaffilmark{3}, 
M. Leventhal\altaffilmark{4}, I. F. Mirabel\altaffilmark{5}, 
L. F. Rodriguez\altaffilmark{6}}

\altaffiltext{1}{Space Sciences Laboratory, University of California Berkeley, 
Berkeley, CA 94720}
\altaffiltext{2}{Center for Astrophysics and Space Sciences, Code 0424, University
of California San Diego, La Jolla, CA 92093}
\altaffiltext{3}{NASA Goddard Space Flight Center, Code 666, Greenbelt,
MD 20771}
\altaffiltext{4}{Dept. of Astronomy, University of Maryland College Park,
College Park, MD 20742}
\altaffiltext{5}{CEN-CEA Saclay, Service d' Astrophysique, 91191 Gif-Sur-Yvette
Cedex, France}
\altaffiltext{6}{Instituto de Astronom\'{i}a, Universidad Nacional Autonoma 
de Mexico, Apdo Postal 70-264, DF04510 Mexico City, Mexico}

\begin{abstract}

We report on long-term observations of the Galactic-bulge black hole
candidates \onee\ and \grs\ with the \it Rossi X-Ray Timing
Explorer\rm. \onee\ has been observed 77 times and \grs\ has been
observed 82 times over the past 1000 days.  The flux of each object
has varied by no more than a factor of 2.5 during this period, and the
indices of the energy spectra have varied by no more than 0.4.  The
power spectra are similar to other black-hole candidates: flat-topped
noise, breaking to a power law.  Each object has exhibited a
brightening that lasted for several months, and we have a found a time
lag between the photon power-law index and the count rate.  In both
sources, the spectrum is softest during the decline from the
brightening.  This behavior can be understood in the context of
thin-disk and advection-dominated accretion flows coexisting over a
wide range of radii, with the implication that both sources have
low-mass companions and accrete via Roche-lobe overflow.

\end{abstract}


\section{Introduction}

\onee\ and \grs\ are by far the brightest persistent sources in the
Galactic bulge above $\sim$50~keV (\markcite{Su91}Sunyaev et
al. 1991).  Their spectra are typical of a black hole low (hard) state
(\markcite{He93}Heindl et al. 1993;\markcite{Su91} Sunyaev et
al. 1991).  Although variable over times of days to years, they spend
most of their time near their brightest observed level.
Both have a core-and-jet structure in the radio
(\markcite{He94}Heindl, Prince, \& Grunsfeld 1994;\markcite{Mir92}
Mirabel et al. 1992; \markcite{Ro92}Rodriguez, Mirabel, \& Mart\'{\i}
1992) and have therefore been described as microquasars.  These
characteristics make this pair of objects a subclass among the black
hole candidates.

     This subclass shares features with other black hole candidates.
Radio jets also appear in the much brighter, and spectacularly
variable objects more usually called microquasars: GRS 1915+105 and
GRO J1655-40 (\markcite{Gr96}Greiner, Morgan \& Remillard
1996;\markcite{Zh97a} Zhang et al. 1997a), whose jets, too, are
brighter and more variable.  Maximum luminosities around $3 \times
10^{37}$ ergs sec$^{-1}$ are shared with Cyg~X-1 and the recently
discovered transient GRS 1737-31 (\markcite{Cu97a}Cui et al. 1997).
The property of being in the hard state at fairly high luminosities
half the time or more is shared only with Cyg~X-1.  The property of
having been observed only in the hard state is shared with GRS
1737-31, GS 2023+338, GRO J0422+32, and GRO J1719-24, although the
total amount of time devoted to these objects varies widely
(\markcite{Zh97}Zhang, Cui, \& Chen 1997; \markcite{Ta95}Tanaka \&
Lewin 1995).
 
Despite the hard spectra of the two objects, there has been some
preliminary evidence of state changes: changes in the spectral shape
of \onee\ above 20 keV from BATSE data (\markcite{Zh97b}Zhang et
al. 1997b), and the detection of weak soft components from \grs\
(\markcite{Me94}Mereghetti, Belloni, \& Goldwurm 1994;
\markcite{He98}Heindl \& Smith 1998; \markcite{Li99}Lin et al. 1999)
and, with marginal significance, from \onee\ (\markcite{He98}Heindl \&
Smith 1998).

Both sources were observed by SIGMA and ART-P to vary between
observations separated by 6 months from a hard X-ray flux of about 130
mcrab (40 mcrab in the 8-20 keV ART-P band) to a level less than 10
mcrab and consistent with zero (\markcite{Cz94}Churazov et al. 1994;
\markcite{Pa94}Pavlinsky et al. 1994).  BATSE has also observed both
\onee\ and \grs\ at a near-zero flux level (\markcite{Zh97b}Zhang et
al. 1997b).  Day-to-day variability is also seen in these data,
including a 1 day jump in ART-P flux from \onee\ from $\sim$3 to
$\sim$18~mcrab.  Both sources show rapid variability with a
flat-topped power spectrum, behavior typical of the hard states of
both black holes and neutron stars (\markcite{Sm97}Smith et al. 1997).

   \onee\ and \grs\ have high Galactic extinction in the optical, and
counterparts have not been identified; only O stars and red
supergiants have been ruled out as companions (\markcite{Ch94}Chen,
Gehrels, \& Leventhal 1994).  There have therefore been no mass
determinations; it has even been suggested (\markcite{Ba91}Bally \&
Leventhal 1991) that \onee\ does not need a companion and could be
accreting directly from a nearby molecular cloud.  However, it has
also been suggested that the lack of a 6.4 keV emission line in the
spectrum of \onee\ strongly constrains the amount of gas immediately 
surrounding the source (\markcite{Ch96}Churazov, Gilfanov, \& Sunyaev 1996).

\section{Observations}

We have observed \onee\ and \grs\ in $\sim$1500 second intervals with
the \it Rossi X-Ray Timing Explorer (RXTE)\rm.  From 1996 February
through 1996 October the observations were spaced one month apart.  We
have observed weekly since 1996 November.  This report is based on data
obtained through 1998 September. Because RXTE cannot point near the Sun, from
late November to late January observations were not taken.  All data
reported here were taken with the Proportional Counter Array (PCA).

The PCA (\markcite{Ja96}Jahoda et al. 1996) consists of five xenon
proportional counters of $\sim$1300~cm$^2$ each, for a total of
6500~cm$^2$, that are sensitive from 2 to 60~keV and share a
1$^{\circ}$ FWHM passively collimated field of view.  We calculate
instrumental background with the standard ``Q6'' model for consistency
in a data set that spans the whole mission.  Although this is not the
most current model, we have found the differences among models to be
negligible for these moderately bright sources.  Our response matrices
are those standard to FTOOLS release 4.1, including the time
dependences of the gain and of the diffusion of xenon into the propane
layer.
 
The pointing directions were offset to avoid other nearby X-ray
sources: A1742-294 and other Galactic Center sources near \onee\ and
GX 5-1 near \grs. The instrumental effective areas resulting from the
offsets were 43\% and 46\% of on-axis values for \onee\ and \grs,
respectively.  Since both sources lie in the Galactic plane,
background observations were made to determine the Galactic diffuse
emission.  For \onee, the background field is opposite in Galactic
longitude and equal in Galactic latitude to the source field.  For
\grs, the background fields are at the same Galactic latitude as the
source field and on either side in Galactic longitude.  The source and
background fields are shown along with some bright sources in the
region in Figure 1.  The coordinates of these pointings are given in
\markcite{Sm97}Smith et al. (1997).

For the spectral analyses, we used only the top layer of the PCA
detectors, in the energy range 2.5 to 25~keV.  In this mode, the
diffuse X-ray background from the Galactic plane is 77 counts
sec$^{-1}$ for \onee\ and 32 counts sec$^{-1}$ for \grs.  The
instrumental background is about 20 counts sec$^{-1}$.  Typical source
count rates for both sources are about 100 counts sec$^{-1}$.
Although \onee\ is somewhat brighter than \grs\ in the 2.5 to 25~keV band, 
it is also more absorbed below a few keV.

All the background-subtracted energy spectra were fitted with an
absorbed power law.  The time histories of the PCA count rate, the
rms variability, and the photon power-law index (PLI) are
shown in Figure 2.  Gaps from November to January of each year are due
to the \it RXTE \rm solar pointing constraint.  A few observations of
each source have been removed due to very high background when the
observation was made immediately after exiting the South Atlantic
Anomaly.  This condition occurred more often for \onee.  After
removing one observation of \onee\ which was contaminated by the new
transient XTE J1739-302 (\markcite{Sm98}Smith et al. 1998), we present
a total of 77 observations of \onee\ and 82 observations of \grs.

\section{Subtle Changes and Hysteresis}

The histories of both \onee\ and \grs\ in Figure 2 clearly show that neither
source turned off during the past 3 years.  The count rate has ranged from
about 60 counts sec$^{-1}$ to 140 counts sec$^{-1}$ in both sources.

This result is in conflict with a recent preliminary report on \grs\ by
\markcite{Co99}Cocchi et al. (1999) using data from the BeppoSAX Wide
Field Camera.  They report that on three occasions (1996 September,
1997 October, and 1998 March) the flux dropped by roughly a factor of
5, becoming so low as to be undetectable.  No such large drops appear
in the \it RXTE \rm data at these times, which are marked by triangles in
Figure 2.  Although neither the BeppoSAX nor the \it RXTE \rm data are
continuous, the \it RXTE \rm dataset has 59 short pointings during the
range of time covered by the 16 short BeppoSAX observations in
\markcite{Co99}Cocchi et al. (1999).  It is therefore highly unlikely
that \it RXTE \rm would miss the large variations reported by BeppoSAX
if they occurred with a random distribution in time.

The \it RXTE \rm spectra show that both sources have remained in the
hard state during the past 3 years, even though the PLI has
occasionally softened slightly.  All the variations discussed below
are subtle changes within the hard state.

Both sources exhibit events of brightening and softening in early
1998 (see Figure 2).  The softening clearly lags the brightening.  By
using the cross-correlation function, we found that there is a
$\sim$58 day lag between PLI and count rate in
\onee\ and a $\sim$36 day lag in \grs.  In both sources, we used only
the data around the peaks in the count rate and PLI for computing
the cross-correlation function.  The peaks occurred between 
22 January 1998 and 11 September 1998.  The brightest
periods are approximately from 2 February 1998 to 28 May 1998 for \onee\
and 22 January 1998 to 12 March 1998 for \grs.

Figure 3 shows scatter plots of PLI vs. count rate for both \grs\ and
\onee\ during these events.  When the points on the scatter plots are
connected, a circle is clearly evident, showing hysteresis between the
two parameters; i.e., the time lags described above can also be
thought of as a phase lag of $\sim90^\circ$.  Because of this hysteresis
effect, our data show that the time lag
could hide real correlations in
scatter plots using data taken over long periods.

We have considered the possibility that these events were
instrumental.  This is unlikely for two reasons: 1) the events are not
simultaneous and 2) they are much greater changes than the known time
evolution of the instrument parameters, such as efficiency and gain.

The brightening/softening event in \onee\ was preceded by a period in which
the PLI was gradually hardening, beginning in March 1997 and
lasting for $\sim$250 days.  In \grs\ there was also a period of 
gradual hardening before the similar event, lasting $\sim$150 days.

\section{Timing Results and Analysis}

The counts for each observation were summed into 31.25 ms bins for the
timing analysis.  The individual observation times ranged from 1000 to
1500 seconds.  Typical power spectra for \grs\ and \onee\ are shown
in Figure 4.  The full energy range of the PCA (2-60 keV) was used for
all the power spectra, but the contribution above 20 keV is small.

We fitted the power spectra with a broken power law (index 0 below the
break and free above it). Typical values of the break frequency range
from 0.1 to 0.8 Hz for both \grs\ and \onee.  Typical values of the
index above the break frequency are around -1. The rms variability integrated 
from 0.004 to 15.8 Hz ranged from 21.5\% to 30.5\% for \grs\ and from 
13.9\% to 28.6\% for \onee.

In these short observations, the statistics were insufficient to
observe any quasi-periodic oscillations (QPOs).  However, in longer
duration observations, QPOs have been observed in both \grs\ and
\onee\ (\markcite{Sm97}Smith et al. 1997).  Figure 5 shows the average
power spectrum of 77 observations for \onee\ and 82 for \grs, and
Figure 6 shows the results from \markcite{Sm97}Smith et al (1997).  
There are no obvious QPOs
in our results.  Since the QPOs exist in the long observations but do
not appear when the short observations are summed, one may conclude
that either the deep observations found a rare appearance of the
QPOs, or that the QPOs drift in frequency with time.
\markcite{Wi98}Wijnands \& van der Klis (1998) have shown a
correlation between the break frequency and QPO frequency in both
neutron stars and black hole candidates.  Since the
break frequencies of \grs\ and \onee\ are variable, then the QPO frequencies
may also be variable.

\section{Energy Spectra}

We fitted the energy spectra of \grs\ and \onee\ with a power law 
absorbed by a column of neutral interstellar gas.  
The variations in PLI were shown in Figure 2.
For \onee, the column depth varied between
$7.4 \times 10^{22}$ and $11 \times 10^{22}$ atoms cm$^{-2}$
and the average value was $9.2 \times 10^{22}$ atoms cm$^{-2}$. 
For \grs, the absorption column varied between $0.71 \times 10^{22}$ 
and $2.3 \times 10^{22}$ atoms cm$^{-2}$ with an average value of 
$1.4 \times 10^{22}$ atoms cm$^{-2}$.
\markcite{Sh96}Sheth et al. (1996), measured the column depth for
\onee\ at $(8.1 \pm 0.1) \times 10^{22}$ atoms cm$^{-2}$ with ASCA, which is
consistent with the range we obtained.  Another ASCA measurement
(\markcite{Me97}Mereghetti et al. 1997) measured the column depth for
\grs\ at $(1.5 \pm 0.1) \times 10^{22}$ atoms cm$^{-2}$, again consistent 
with our range.

The column depth is mostly determined by the spectral shape below
$\sim$4~keV, and the PLI mostly by the higher-energy part of the
spectrum.  At the lowest energies, we are most vulnerable to
uncertainties about the influence of diffuse emission and soft sources
on the edges of both fields of view (see \S 2).  We therefore do not
claim that the range of absorption columns obtained is evidence of
real variability. There is no correlation between the column depth
and PLI in either source, and we are confident that the variations
in PLI are real.

The PLI ranged from 1.37 to 1.76 for \onee\, and from 1.45 to 1.86 for
\grs.  The values found by \markcite{He98}Heindl and Smith (1998) for
the PLI of \grs\ and \onee, using
their deep pointings to both sources in August and March of 1996,
were 1.54 and 1.53, respectively.  These
values included HEXTE data and were derived with a model which
included an exponential cutoff at high energies.  With the simpler
power-law model used here, the indices from the monitoring
observations just before and after each deep pointing average to 1.67
and 1.68 for \grs\ and \onee, respectively.  This is consistent with
the expectation that, in the absence of a cutoff in the model, the
effect of the cutoff will appear in a softening of the fitted PLI.
The statistics in individual monitoring observations are not good
enough to measure the cutoff and PLI independently.

\section{Comparison to Cygnus X-1}

The similarities between Cyg~X-1 and \onee\ and \grs\ suggest that
these three sources are similar objects.  Some of these similarities
were mentioned in \S 1: the x-ray luminosities, hard spectra, and
persistent activity of all three sources. Another similarity is the
shape of the power spectra.  When in the hard state, all three sources
show white noise up to $\sim$.5~Hz (see \S 4) and break to a power
law, with an index above the break of $\sim$ -1.  Cyg~X-1 displays a
relationship between the break frequency and the rms variability
integrated over frequency.  This
behavior was first illustrated by \markcite{Be90}Belloni and Hasinger
(1990).  They showed that the low-state power spectrum for Cyg~X-1
always had the same normalization above the break frequency, which
varied.  \markcite{Miy92}Miyamoto et al. (1992) noted that this held
true from one black hole candidate to another.  We searched for the
same effect in \grs\ and \onee.  We divided our data into three
groups, those with the largest, smallest, and near-average rms and
averaged the power spectra in each group. Because of this averaging,
the break frequencies are more rounded than Belloni \& Hasinger showed
for Cyg X-1, but otherwise Figure 7 shows that \grs\ and \onee\ are
similar to Cyg~X-1 in this respect.

Unlike \onee\ and \grs, Cyg~X-1 has been observed in a true soft
state, in which a soft thermal component was dominant
(e.g. \markcite{Cu97b}Cui et. al. 1997b).  When Cyg~X-1 was in the
soft state, the PLI was -2.2.  A similar index was seen by BATSE above
20 keV in \onee\ while that source was faint in the BATSE band
(\markcite{Zh97b}Zhang et al. 1997b).
Simultaneous observations at lower energies during another occurrence
of this state are needed to confirm that it is a soft state similar to
that in Cyg~X-1 and other black-hole candidates.
  
\section{Discussion}

\subsection{Dynamical model for hysteresis}

We can qualitatively explain the hysteresis or time lag between
brightening and softening in 1E 1740.7-2942 and GRS 1758-258 in the
context of some recent models of black-hole accretion.  These models
(e.g. \markcite{Ch95}Chakrabarti \& Titarchuk 1995, 
\markcite{Es98}Esin et al. 1998) have two
components in the outer regions of the flow: a standard Keplerian
disk, physically thin and optically thick,
and an optically thin, physically thick halo or corona.  The mass in
the halo is nearly in radial free-fall, and it advects most of its
accretion energy into the black hole rather than radiating it as the
Keplerian disk does (\markcite{Ic77}Ichimaru 1977; 
\markcite{Re82}Rees et al. 1982;
\markcite{Na95}Narayan \& Yi 1995;
\markcite{Ab95}Abramowicz et al. 1995).  

In early disk-plus-corona models, the corona was produced locally by
the Keplerian disk, and did not accrete independently and advectively
(e.g.  \markcite{Li77}Liang \& Price 1977,
\markcite{Bi77}Bisnovatyi-Kogan \& Blinnikov 1977).  In the newer
models, it is an equally valid and independent solution of the
hydrodynamic equations.

Within a certain radius, the Keplerian disk is unstable, and only a
very hot solution remains (an advection-dominated flow in the model of
\markcite{Es98}Esin et al. 1998, and a shocked flow in the model of
\markcite{Ch95}Chakrabarti \& Titarchuk (1995)).  The soft, thermal
component of black-hole-candidate spectra is attributed to the inner
part of the Keplerian disk, near this boundary.  The hard, power-law
component is attributed to inverse Comptonization of these soft
photons in the very hot inner parts of the advective flow.

While matter in the advective flow is nearly in free-fall, matter in
the thin disk only accretes after a gradual loss of angular momentum
via viscous torques.  The timescale for this process is approximately
(\markcite{Fr92}Frank, King \& Raine 1992)
\begin{equation}
t_{\rm{visc}} \; \sim \; 3 \times 10^{5} \alpha ^{-4/5} 
\left(\frac{\dot{M}}{10^{16} \rm{g/s}}\right)^{-3/10} 
\left(\frac{M}{M_{\sun}}\right)^{1/4} 
\left(\frac{R}{10^{10} \rm{cm}}\right)^{5/4} \;
\rm{s},
\end{equation}
where $M$ is the black-hole mass, $\dot{M}$ the accretion rate,
$R$ the disk radius, and $\alpha$ the viscosity parameter
$(0 < \alpha \lesssim 1)$.  
\markcite{Ch95}Chakrabarti \& Titarchuk (1995) pointed
out that if the mass accretion rate were increased at the outer edge
of both flows simultaneously, it would arrive at the central regions
of the advective flow first.  Thus the hard component would brighten
first, with the soft component only brightening after a delay
approximately equal to the viscous time.  We offer this delay as one
explanation for the hysteresis we observe between brightening and
softening.  If we use $M = 10M_{\sun}$, 
$\dot{M} = 10^{17}$g/s, $t_{\rm{visc}}$ equal to the measured delays
(see \S 3), and $\alpha$ = 0.3 
(\markcite{Es98}Esin et al. 1998) to solve
equation (1) for $R$, we find $R = 5 \times 10^{10}$ cm 
($3 \times 10^4 \; GM/c^2$) for \onee\ and 
$R = 3 \times 10^{10}$ cm ($2 \times 10^4 \; GM/c^2$) for \grs.
These disk sizes are typical of low-mass
x-ray binaries accreting by Roche-lobe overflow, and are larger
than the disks expected in systems accreting winds from massive
companions (\markcite{Fr92}Frank et al. 1992).

\subsection{Disk-evaporation model for hysteresis}

An alternative explanation for the lag between brightening and
softening is quasi-static rather than dynamic: the flows can be
allowed to reach an equilibrium configuration after every infinitesimal
increase in accretion rate.  This explanation relies on a
characteristic common to the models of \markcite{Es98}Esin et
al. (1998) and \markcite{Ch95}Chakrabarti \& Titarchuk (1995): as the
accretion rate rises, the inner edge of the Keplerian disk moves
inwards.  If this edge were sufficiently far out to begin with, the
spectral changes due to its inward advance would at first be
restricted to the EUV and soft x-ray ranges, which are not observable
for sources deep in the Galactic bulge.  The only effect on the hard
x-rays of increasing the accretion rate would be a brightening.
Eventually, the disk would move in so far that it would begin
to replace the hard-x-ray-emitting region of the advective or shocked
flow, resulting in a spectral softening.

If the response to the reduction in accretion rate back to the normal
level were equally quasi-static, one would expect to see a bright and
hard phase on the decline, i.e. the peaks in Figure 2 would be
symmetric in time.  However, it may be that the thin disk, once
established at smaller radii, takes a significant amount of time to
evaporate.

It has been noted that there is hysteresis in the hard/soft/hard
transitions of soft x-ray transients (\markcite{Mi95}Miyamoto et
al. 1995).  In a typical outburst of this class of black-hole
candidate, the system remains in the hard state as the luminosity
rises quickly from quiescence to near maximum, then switches to the
soft state, then returns to the hard state only when the luminosity is
of order 1\% of maximum.  The quick rise in accretion rate and
quick transition to the soft state are thought to be due to the
rapid propagation of a thermal-ionization instability in the disk
(\markcite{Ca85}Cannizzo et al. 1985).  This mechanism is not relevant to 
\onee\ and \grs, since their usual accretion rates are high enough
that the disk would remain ionized by the x-rays from the central regions
of the accretion flow.  

The return of the transients to the hard state, however, may be
relevant to our observations.  \markcite{Mi96}Mineshige (1996)
interpreted this return as a transition from a Keplerian disk to an
advective flow.  Both the disk and advective flows are stable over
most of the accretion rates traversed during the decline, but the
transition doesn't take place until the disk solution becomes unstable
at very low accretion rates.  In other words, the disk, once it is
established, tends to persist in regimes where both solutions are
stable.

The typical luminosity of both \onee\ and \grs\ is $2 \times 10^{37}$
erg s$^{-1}$ from 1-200 keV (\markcite{He98}Heindl \& Smith 1998).
Our data never deviate by more than about 50\% from this value.  This
is roughly 1-7\% of Eddington luminosity for black holes of 3-20 $M_{\sun}$,
and is orders of magnitude higher than the luminosity where
the idealized transient of \markcite{Mi96}Mineshige (1996) is forced
to return to the hard state.  Nonetheless, if the added regions of
inner disk in the transients persist for a month or more at accretion
rates where the advective flow would also be stable, then the more
modest inward extensions of the disk which occur when \onee\ and \grs\
brighten might persist as long.  The time asymmetry in our data, in
this interpretation, would be due to the evaporation time of the inner
disk being longer than the time in which the accretion rate returns to
normal.

\section{Conclusions}

We have presented the most detailed long-term coverage of these black
hole candidates to date.  In this 3 year period, we have never seen
either source at a flux level less than half its maximum.  Although 
neither source has entered the soft (high) state in this time,
we have seen variations in spectral index within the range usually
associated with the hard or low state (photon PLI $< 2.0$).

There is hysteresis when \grs\ and \onee\ brighten and soften within
the hard state, with the softening lagging the brightening by 1-2
months.  This hysteresis could be due to the different propagation
times in a disk and halo of an increase in $\dot{M}$ (\S 7.1), 
or else to a persistence of the thin disk after $\dot{M}$ returns to
normal (\S 7.2).  If the former is the correct interpretation,
the lag time implies accretion disks of the size usually associated
with accretion from a low-mass companion overflowing its Roche lobe.

We find that the weekly observations of \grs\ and \onee\ reveal no
QPOs when summed up over many weeks.  This leads us to the conclusion
that either our deep observations observed rare appearances of the
QPOs, or, more likely, that they drift in frequency with time,
consistent with the behavior described by \markcite{Wi98}Wijnands \&
van der Klis (1998) for other sources.  Both objects show the same
relationship between the break frequency of the power spectrum and the
total rms variability as other hard-state sources.

\acknowledgements

We would like to thank Ann Esin, Lars Bildsten, and Lev Titarchuk for
useful discussions on the interpretation of the hysteresis results.
This work was supported by NASA grants NAG5-4110, NAG5-7522, and
NAG5-7265.

\newpage

\begin{figure}[t!]
\centerline{\epsfysize=6.5in \epsfbox{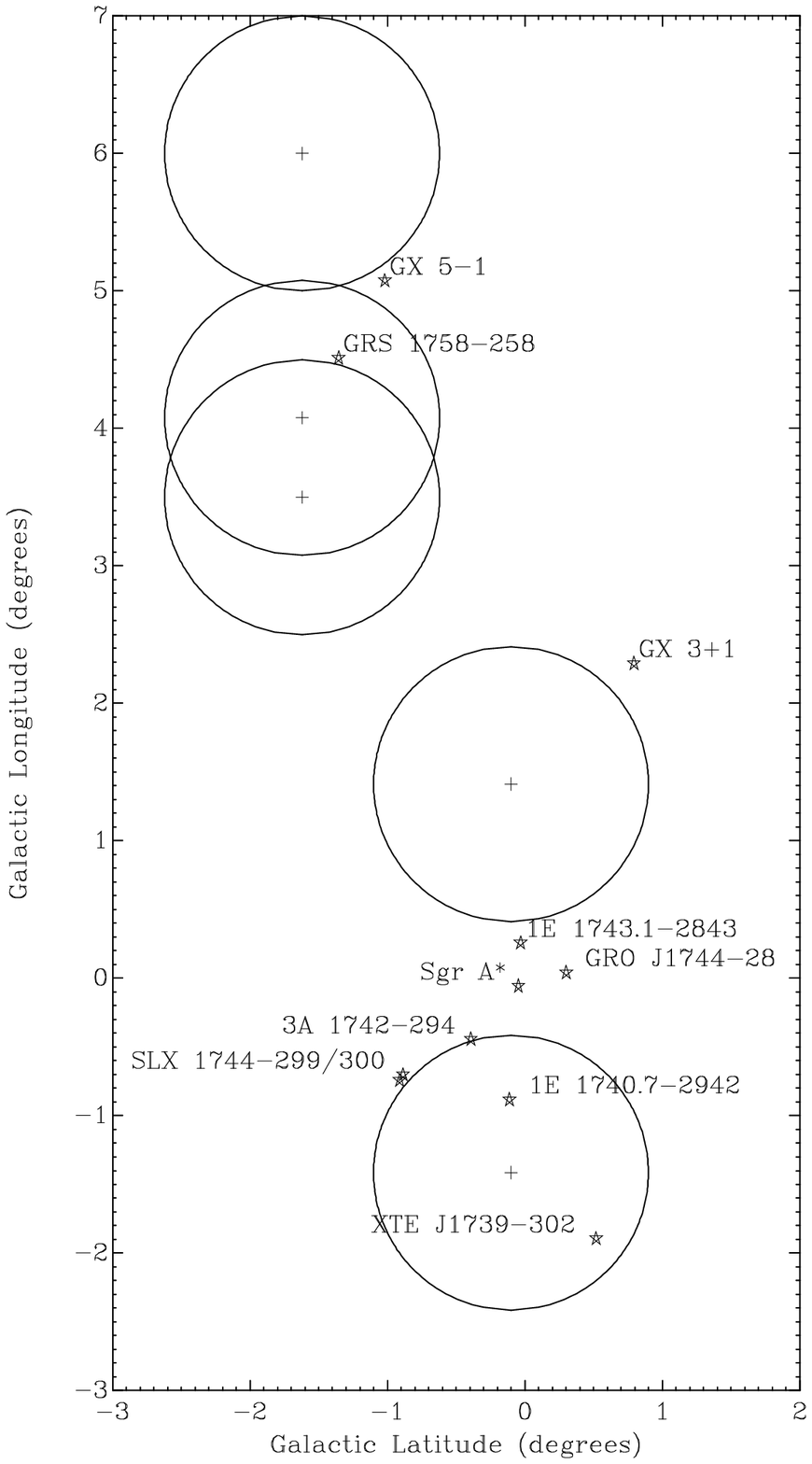}}
\caption{
The Galactic Center region showing bright sources and the \it RXTE \rm
pointings used for this work.  The circles shown have a radius of
$1^{\circ}$, near the 0\% response contour for both the PCA and HEXTE
instruments. The \onee\ background pointing is above the \onee\
source pointing on this plot; the \grs\ pointing and its backgrounds
on either side are clustered near the top.}
\end{figure}

\newpage

\begin{figure}[t!]
\centerline{\epsfysize=6.5in \epsfbox{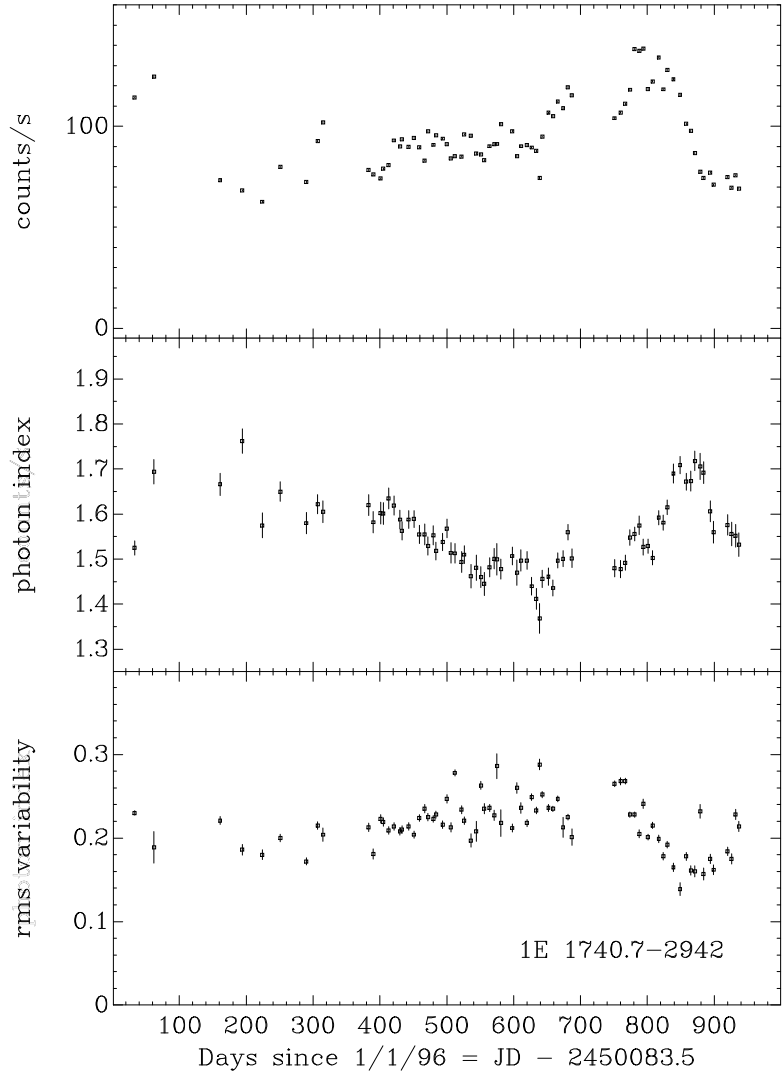}}
\caption{ For caption, see Figure 2b, next page.}
\end{figure}

\setcounter{figure}{1}

\begin{figure}[t!]
\centerline{\epsfysize=6.5in \epsfbox{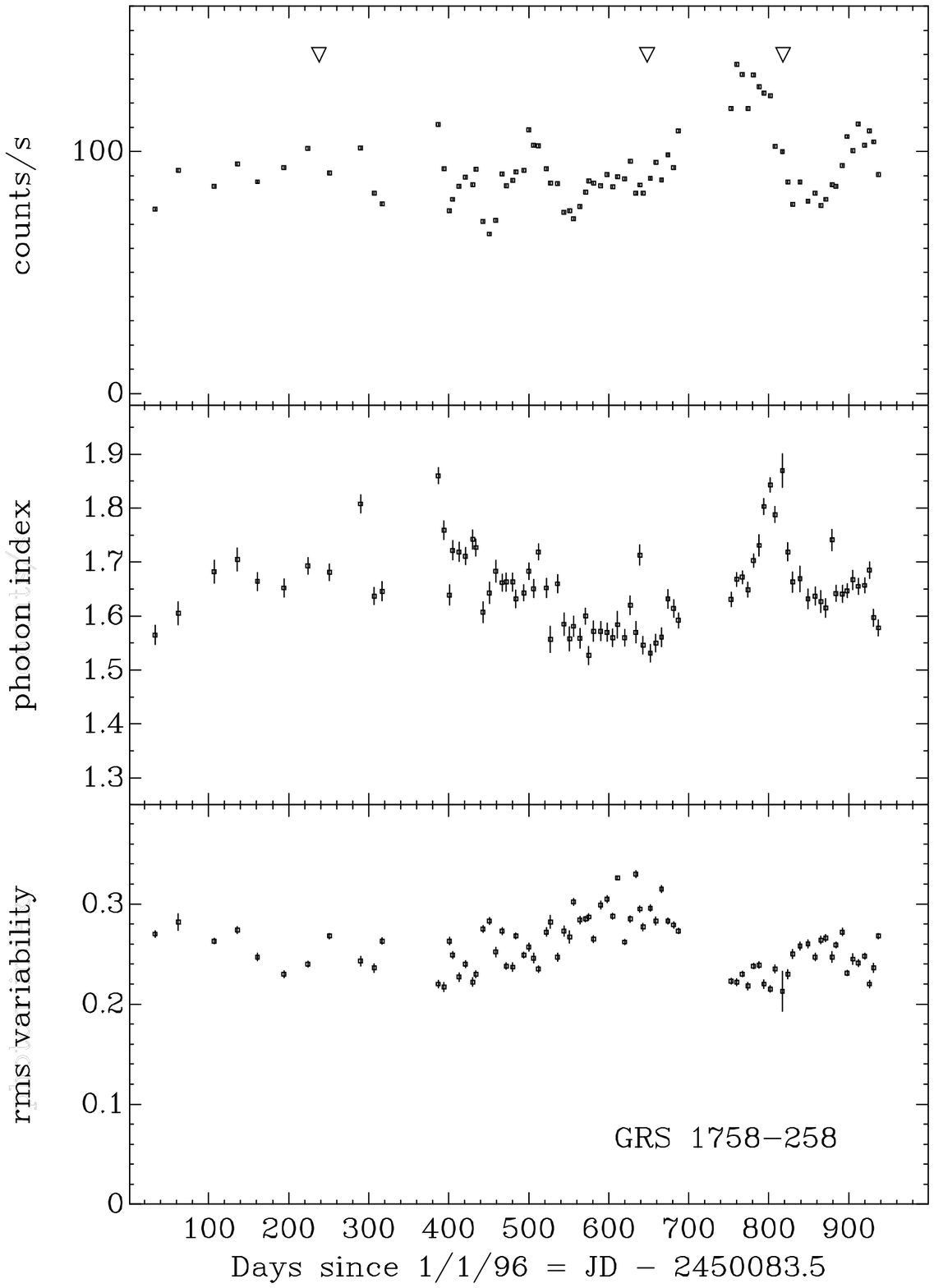}}
\caption{ 
Time histories of count rate, PLI, and
fractional rms variability integrated over frequency for \onee\ and
\grs.  In each source, the count rate and the PLI each peak at the
beginning of 1998, with the PLI lagging the count rate.  The
approximate dates that BeppoSax observed drops to near zero flux in
\grs\ are marked with triangles (see \S 3); the drops are not seen
here.}
\end{figure}

\begin{figure}[t!]
\centerline{\epsfysize=3.2in \epsfbox{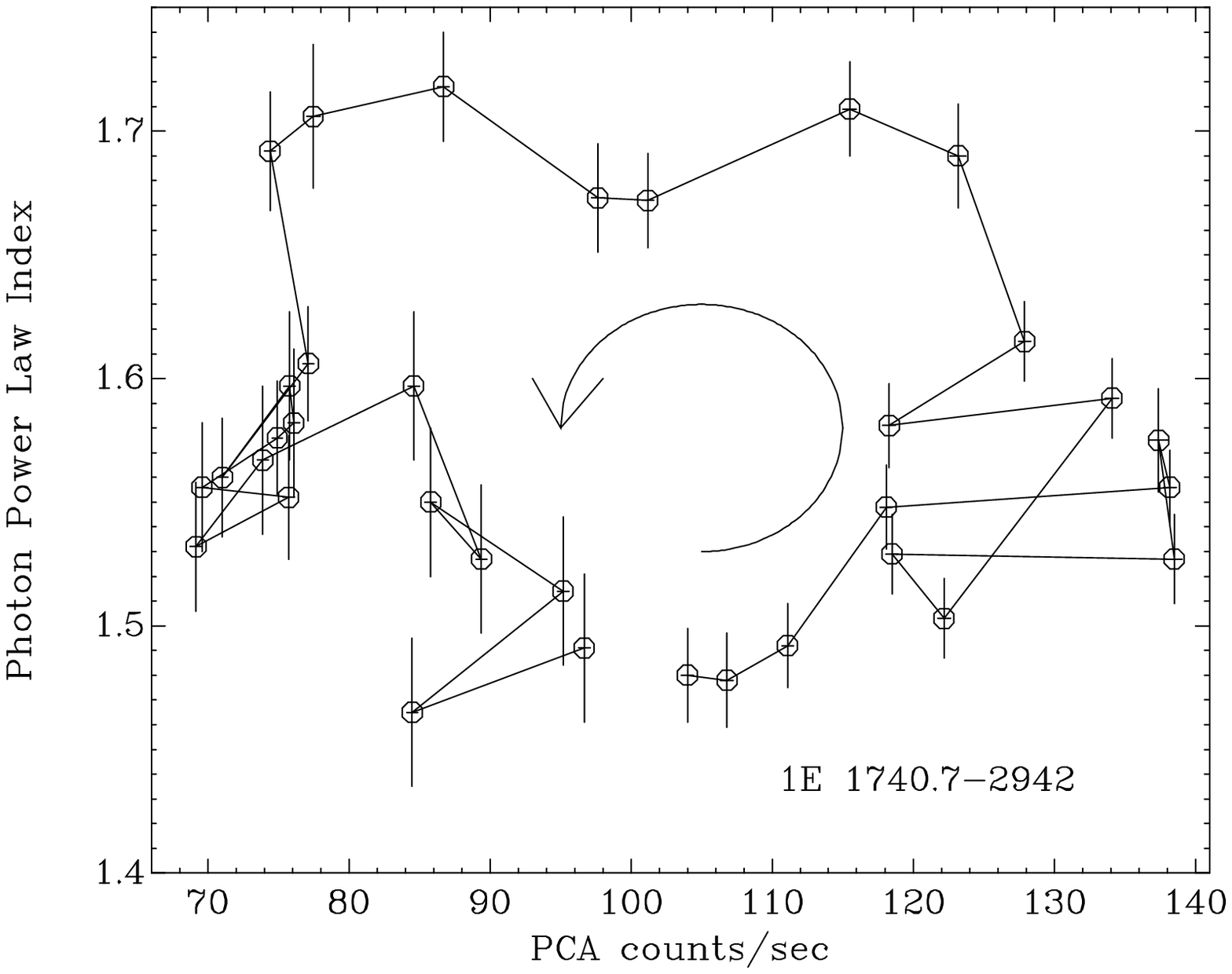}}
\centerline{\epsfysize=3.2in \epsfbox{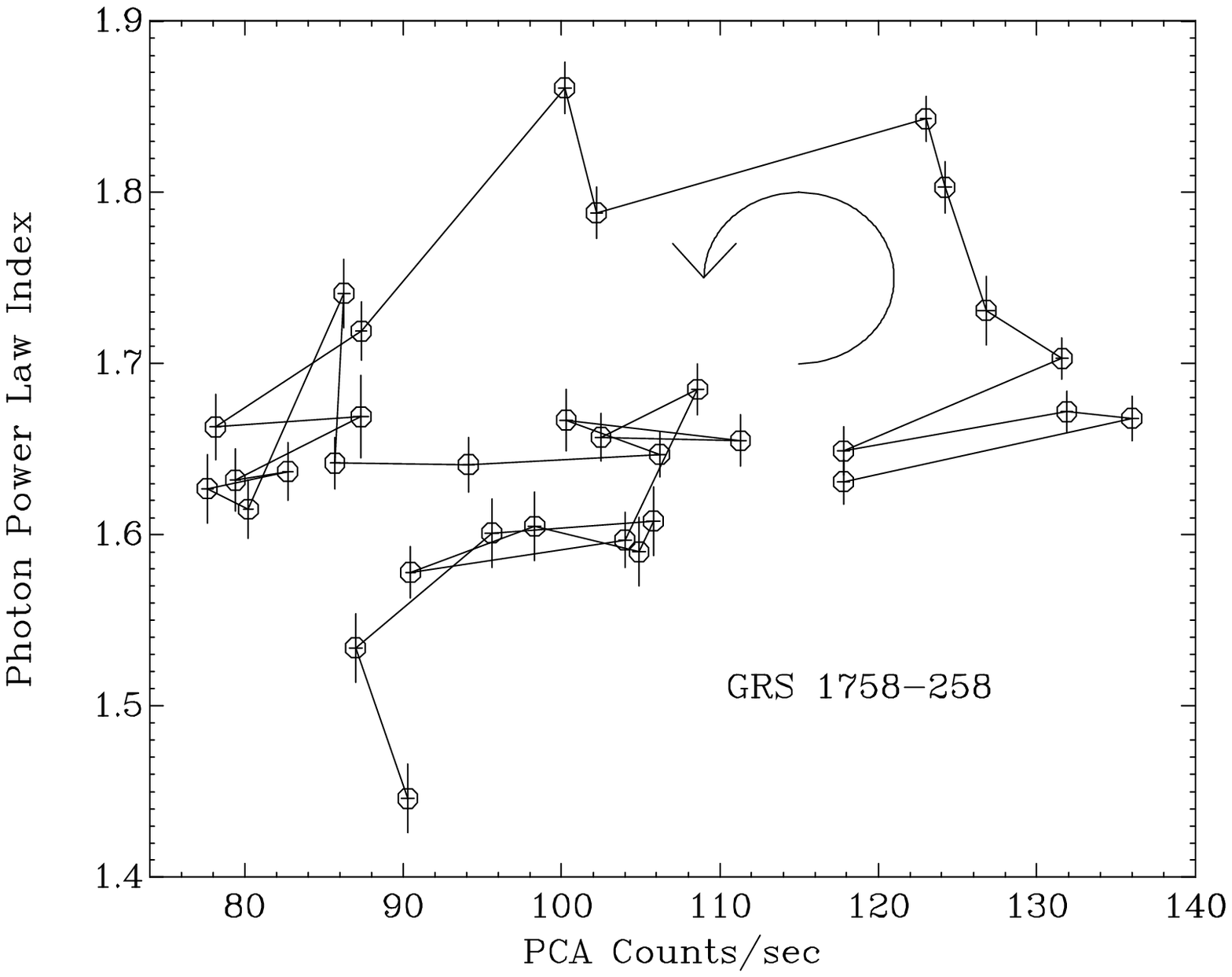}}
\caption{
Hysteresis in \onee\ and \grs.  The arrow shows the sequence of the 
sources' evolution.  For both the PCA count rate and PLI, only the data
around the peaks were used. }
\end{figure}

\begin{figure}[t!]
\centerline{\epsfysize=3.2in \epsfbox{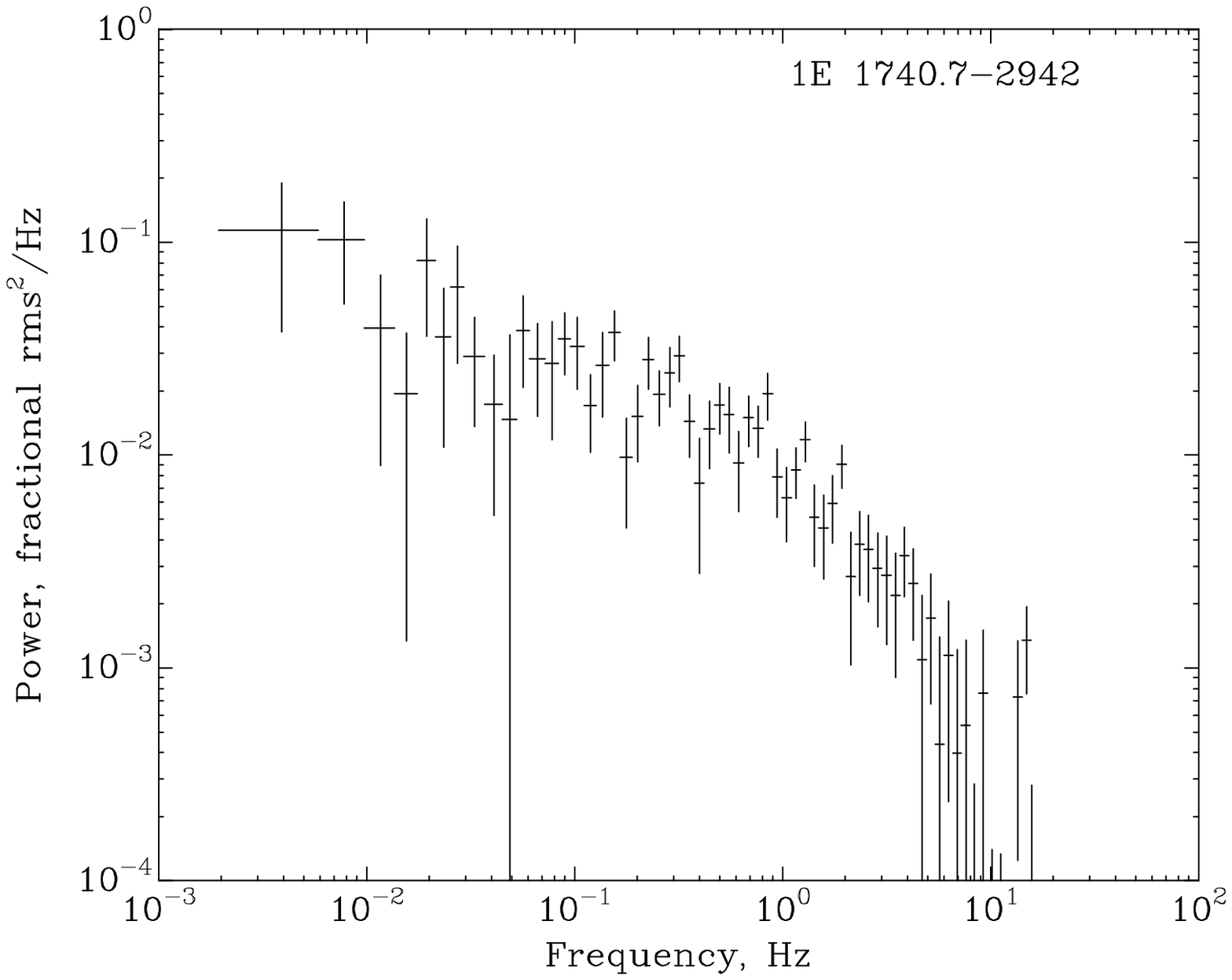}}
\centerline{\epsfysize=3.2in \epsfbox{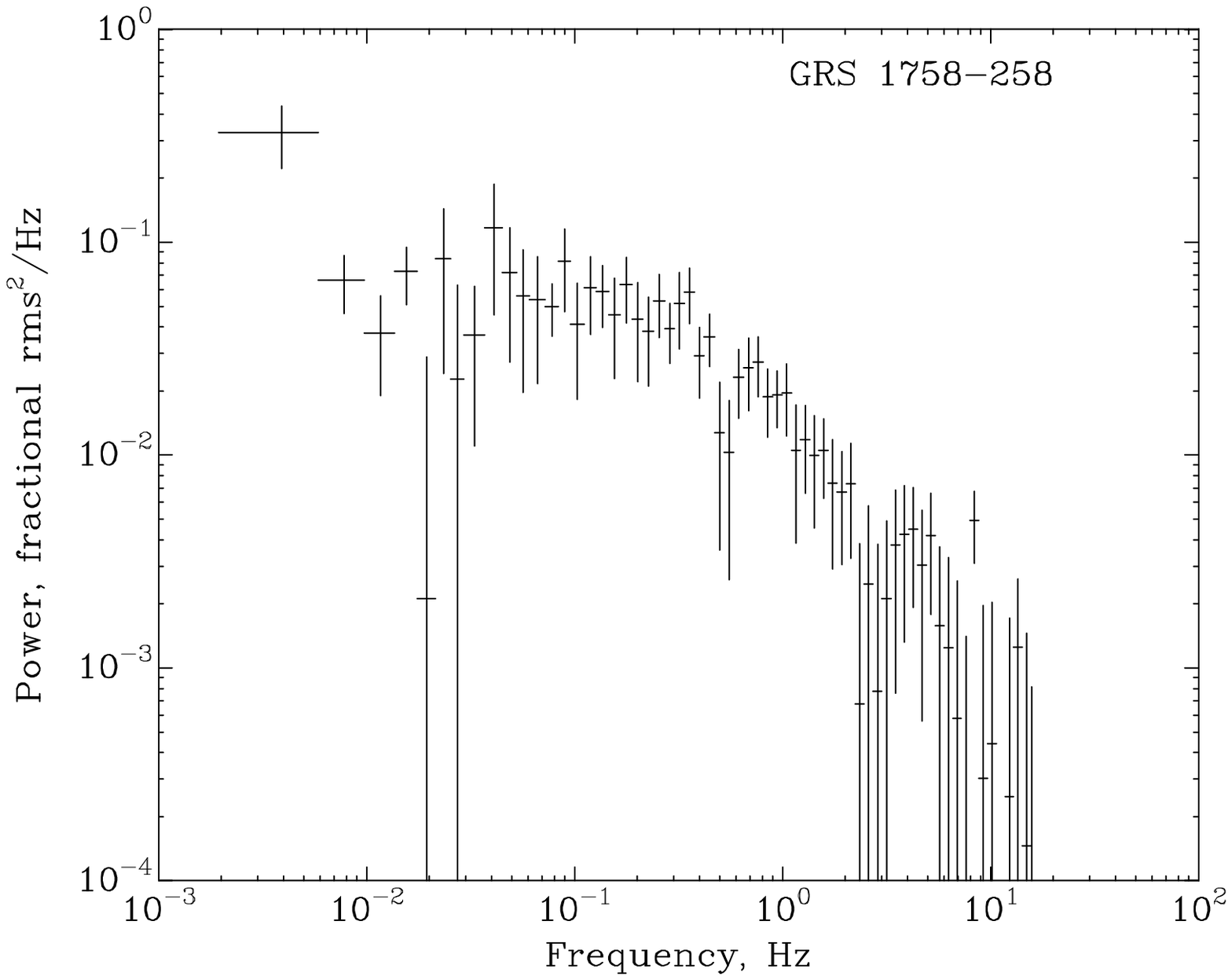}}
\caption{
Typical power spectra for one 1500 second observation
of \onee\ and \grs.  It is apparent that the statistics are not
good enough to observe any QPOs of the sort shown in figure 5.
Power due to Poisson statistics has been subtracted. }
\end{figure}

\begin{figure}[t!]
\centerline{\epsfysize=3.2in \epsfbox{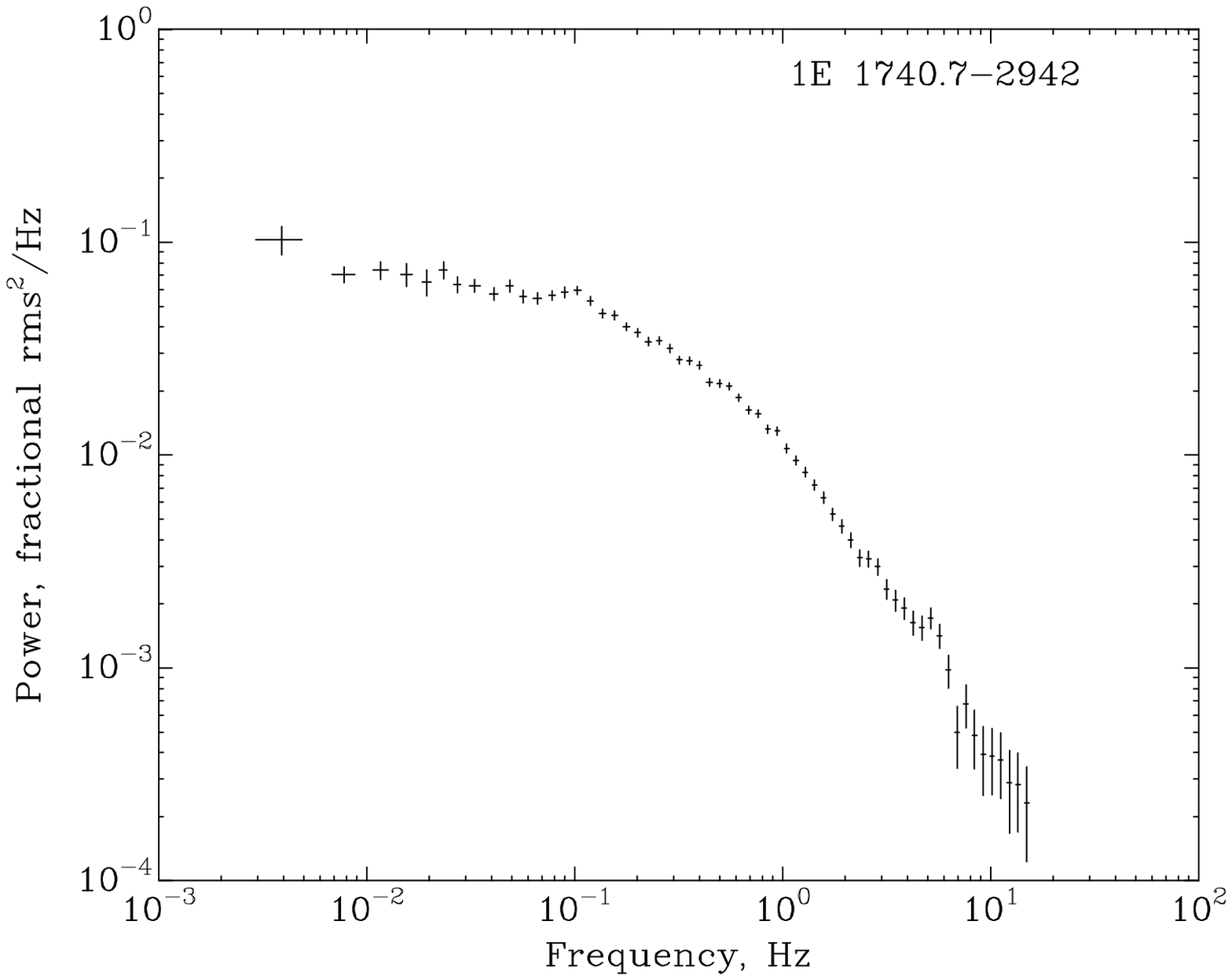}}
\centerline{\epsfysize=3.2in \epsfbox{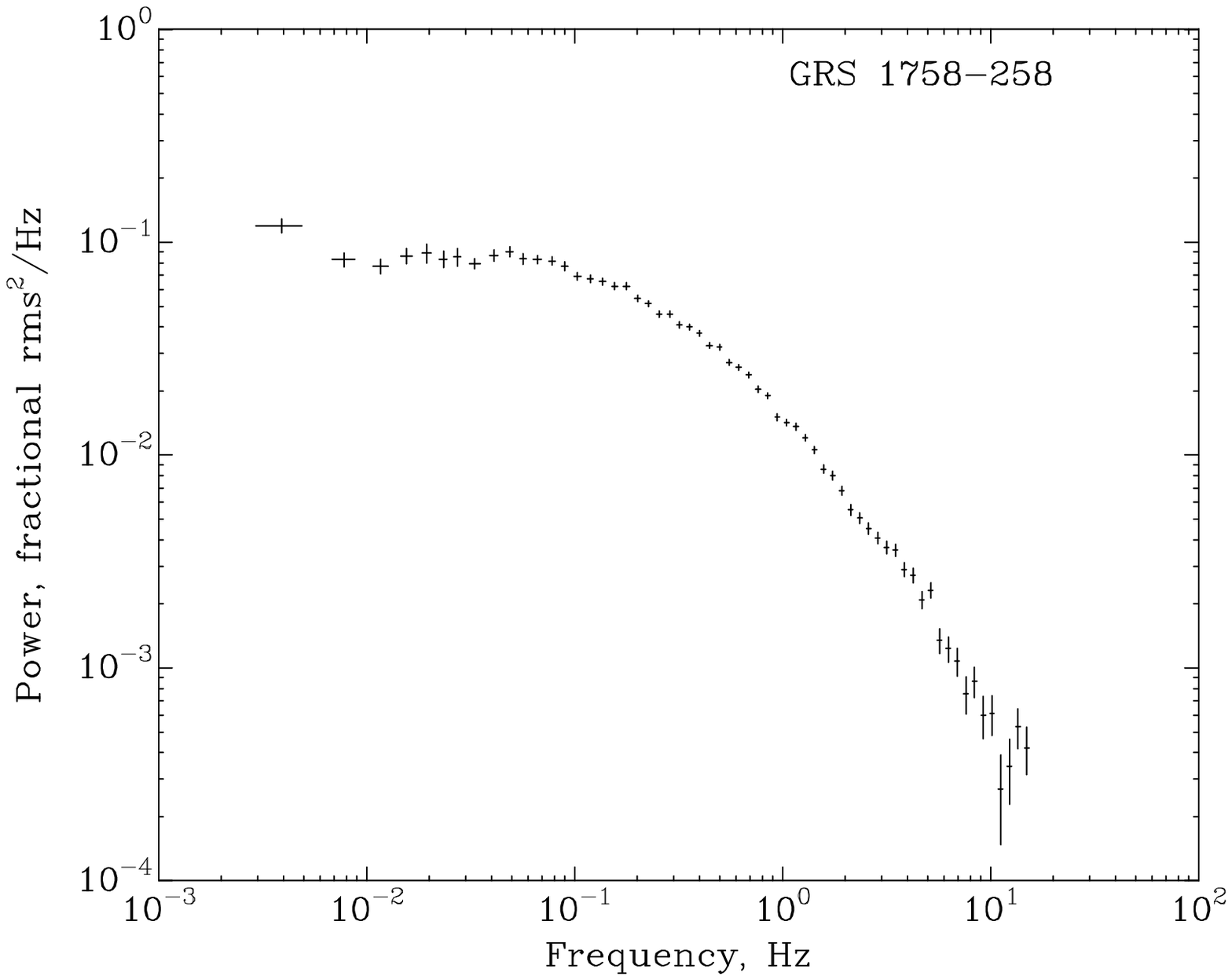}}
\caption{
Average power spectrum of 77 observations for \onee\ and 
82 observations for \grs.   }
\end{figure}

\begin{figure}[t!]
\centerline{\epsfysize=3.2in \epsfbox{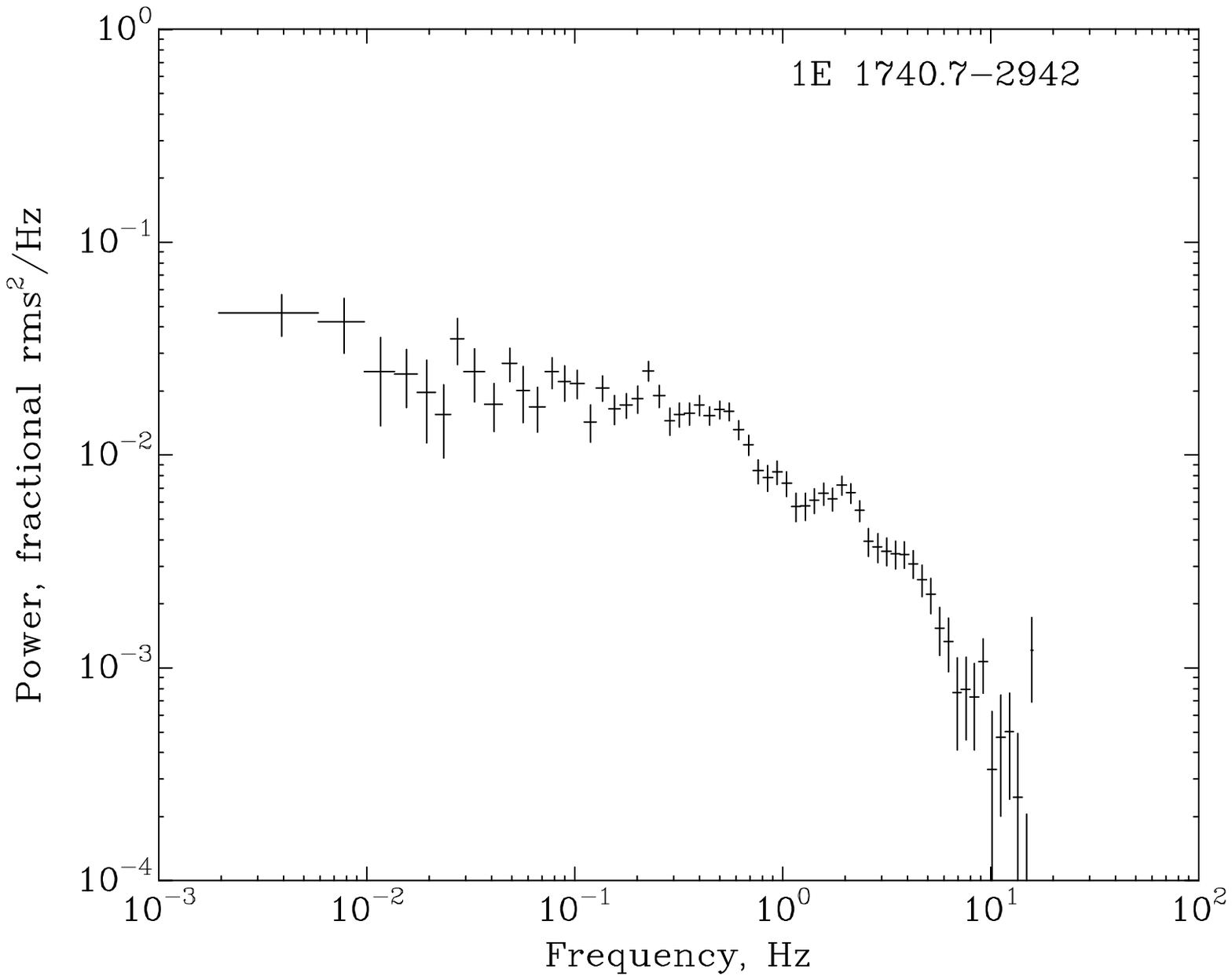}}
\centerline{\epsfysize=3.2in \epsfbox{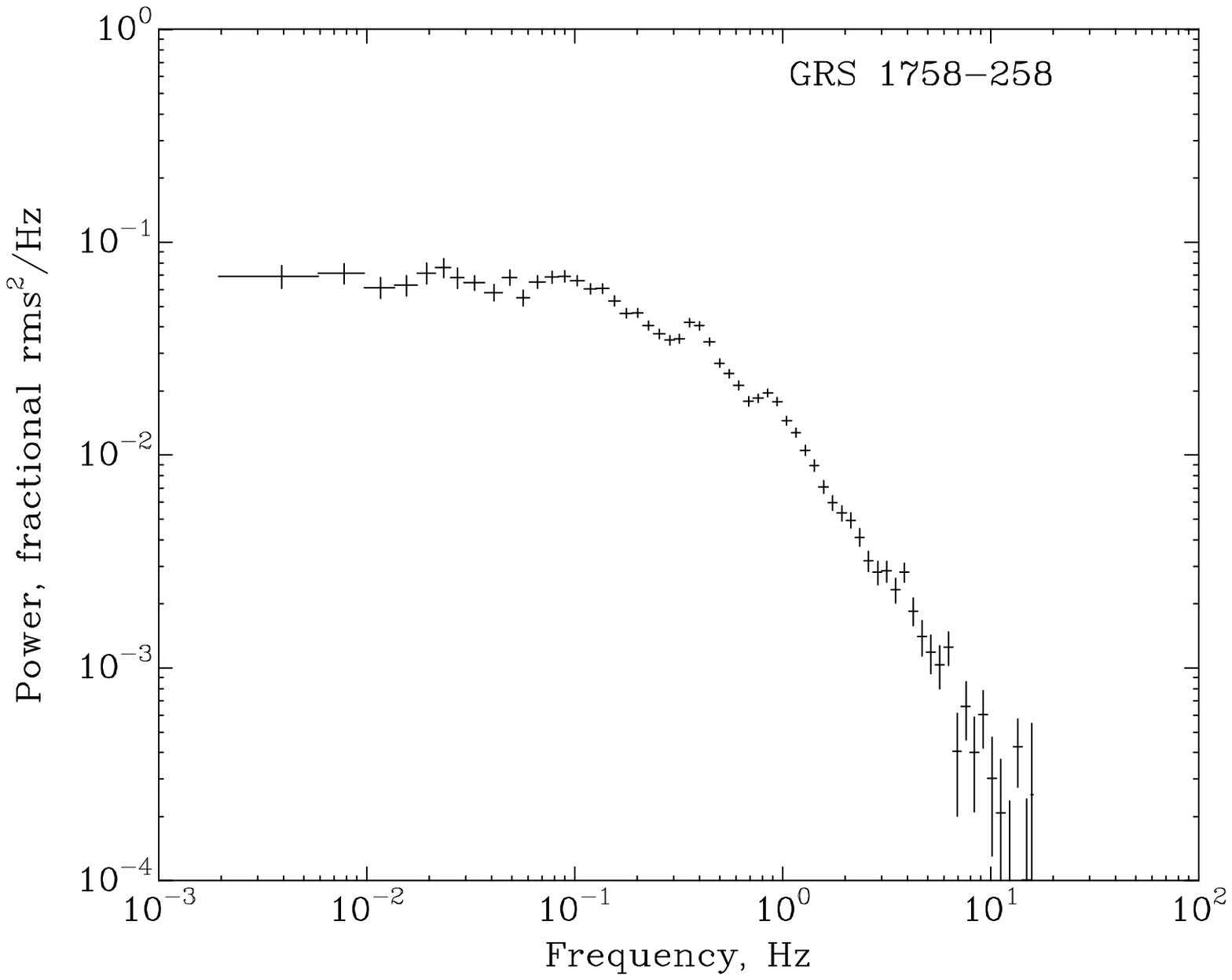}}
\caption{
Power spectra from deep observations of \onee\ and \grs\
(Smith et al. 1997).  Note the QPO pairs in both black hole 
candidates.  These QPOs are absent in the summed weekly 
observations (Figure 4).  }
\end{figure}

\begin{figure}[t!]
\centerline{\epsfysize=3.2in \epsfbox{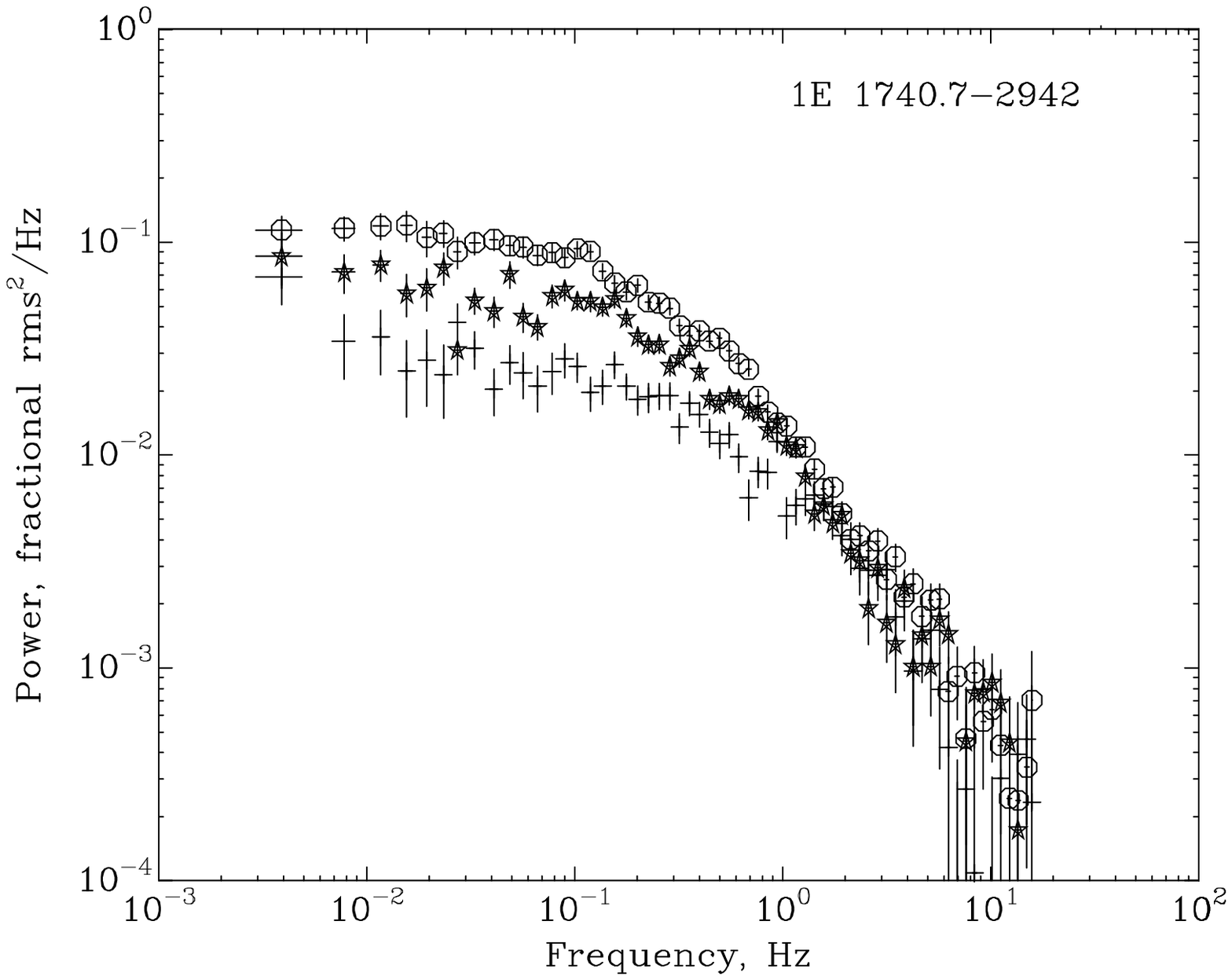}}
\centerline{\epsfysize=3.2in \epsfbox{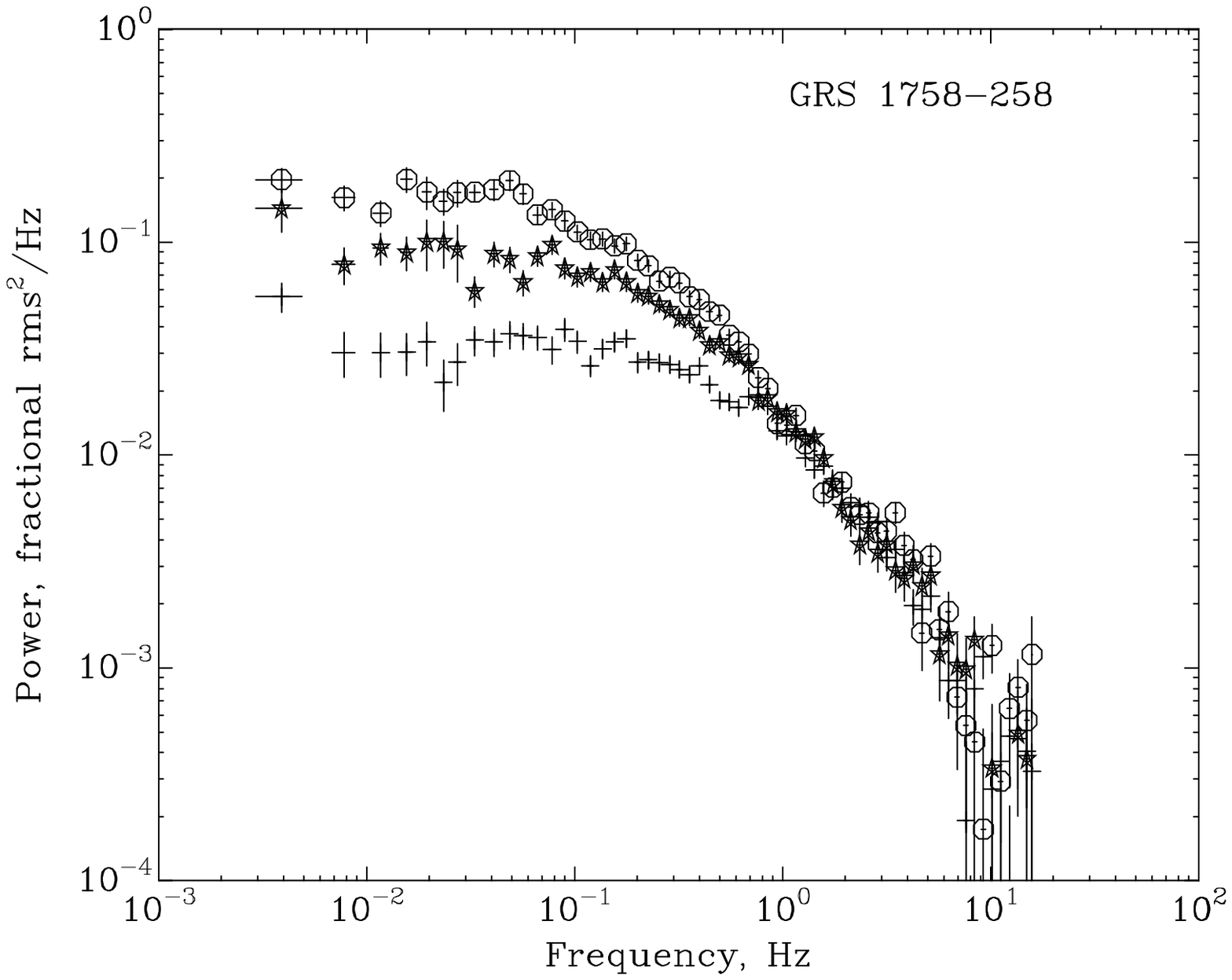}}
\caption{
Belloni and Hasinger (1990) first found a relationship
between power spectrum break frequency and integrated rms 
variability in Cyg~X-1. 
This same relationship is displayed in \onee\ and \grs: the normalization is
fixed above the break frequency.  The power spectra with the highest 
frequency-integrated rms were 
averaged to form the top curve, the spectra with the middle rms were averaged
to form the middle curve, and the spectra with the lowest rms were averaged to 
form the bottom curve.  }
\end{figure}

\end{document}